\documentclass[prb,twocolumn,preprintnumbers,amsmath,amssymb,floatfix]{revtex4-1}

\usepackage{graphicx}
\usepackage{amssymb}
\usepackage{amsmath}
\usepackage{dcolumn}
\usepackage{bm}
\usepackage{color}

\makeatletter 
\def\@hangfrom@section#1#2#3{\@hangfrom{#1#2#3}}
\makeatother

\makeatletter        
\def\@biblabel#1{[#1]}
\makeatother

\newcommand{\cmt}[1]{} 

\usepackage[normalem]{ulem}
\usepackage{xcolor}
\renewcommand\sout{\bgroup\markoverwith{\textcolor{red}{\rule[0.5ex]{2pt}{0.4pt}}}\ULon}

\mathchardef\mhyphen="2D

\usepackage{threeparttable}
\usepackage{hyperref}
\begin{document}
\title{Natural liquid organic hydrogen carrier with low dehydrogenation energy: A first principles study}

\author{Chunguang Tang$^{1,2,*}$}
\email{chunguang.tang@anu.edu.au}
\author{Shunxin Fei$^3$}
\author{G. David Lin$^{1}$}
\author{Yun Liu$^{1,2*}$}
\email{yun.liu@anu.edu.au}
\affiliation{$^1$ Research School of Chemistry, The Australian National University, Canberra, Australia;\\ $^2$ Energy Change Institute, The Australian National University, Canberra, Australia;\\ $^3$ School of Materials Science and Engineering, Anhui University of Technology, Maanshan 243002, People's Republic of China}

\begin{abstract}
Liquid organic hydrogen carriers (LOHCs) represent a promising approach for hydrogen storage due to their favorable properties including stability and compatibility with the existing infrastructure. However, fossil-based LOHC molecules are not green or sustainable. Here we examined the possibility of using norbelladine and trisphaeridine, two typical structures of Amaryllidaceae alkaloids, as the LOHCs from the sustainable and renewable sources of natural products. Our first principles thermodynamics calculations reveal low reversibility for the reaction of norbelladine to/from perhydro-norbelladine because of the existence of stabler isomers of perhydro-norbelladine. On the other hand, trisphaeridine is found promising due to its high hydrogen storage capacity ($\sim$5.9 wt\%) and favorable energetics. Dehydrogenation of perhydro-trisphaeridine has an average standard enthalpy change of $\sim$54 KJ/mol-H$_2$, similar to that of perhydro-\textit{N}-ethylcarbazole, a typical LOHC known for its low dehydrogenation enthalpy. This work is a first exploration of Amaryllidaceae alkaloids for hydrogen storage and the results demonstrate, more generally, the potential of bio-based molecules as a new sustainable resource for future large-scale hydrogen storage.

 \vspace{10pt}

Published at \href{https://doi.org/10.1016/j.ijhydene.2020.08.143}{International Journal of Hydrogen Energy, 45, 32089 (2020).}
\end{abstract}
\maketitle

\section{Introduction}
Driven by the need to reduce carbon footprint, hydrogen has attracted extensive research attention as an ideal energy carrier. The storage and transport of hydrogen via traditional techniques using compressed gas and liquefaction have great cost, efficiency, and safety concerns. Alternative methods of absorbing hydrogen onto carbon based materials or alloys are limited by the storage capacity in terms of weight and volume \cite{felderhoff_hydrogen_2007}. Recently, liquid organic hydrogen carriers (LOHCs) have emerged as a potential alternative storage approach that has attracted extensive research interests, as discussed in a number of reviews \cite{sotoodeh_overview_2013, preuster_liquid_2017, modisha_prospect_2019}. A LOHC system can be hydrogenated  and dehydrogenated in a reversible process and the hydrogen-lean form of the carrier can be recycled. The stability of hydrogenated LOHCs originating from the chemical bonding with hydrogen makes them ideal for distant transport of hydrogen.


An ideal LOHC features a combination of various properties including hydrogen storage capacity, stability, reaction rate, cost, safety, and compatibility with the existing technology and facilities \cite{modisha_prospect_2019}. Among these properties, the enthalpy of dehydrogenation, $\Delta H^{\rm{dh}}$, is an important parameter, which defines the energy cost for the reaction and is related to the necessary dehydrogenation temperature. For a given application, an ideal LOHC should have $\Delta H^{\rm{dh}}$ such that the Gibbs free energy change for the reaction is near zero at the operating temperature \cite{pez_hydrogen_2006}. For example, the ideal $\Delta H^{\rm{dh}}$ range for LOHC used for a proton exchange membrane fuel cell, which operates at relatively low temperatures ($\sim$80 $^\circ$C), is 42-54 kJ/mol-H$_2$ \cite{modisha_prospect_2019}. For polycyclic hydrocarbons $\Delta H^{\rm{dh}}$ generally decreases with an increasing number of aromatic sextets or with some carbon atoms being replaced by nitrogen or oxygen heteroatoms \cite{pez_hydrogen_2006}. Nevertheless, increasing the number of aromatic rings beyond certain optimum limits could result in other constraints, such as difficulty in liquefaction of the molecules \cite{pez_hydrogen_2006}, and the choice of LOHC is a compromise between these competing factors.

Extensive research efforts have been carried out for the search of LOHCs \cite{kariya_efficient_2002, kariya_efficient_2003,suttisawat_microwave_2012,mueller_liquid_2015,yang_study_2018}. For example, dehydrogenation reactions of cyclohexane, methylcyclohexane, tetralin, and decalin were studied using Pt and Pt-M (M = Re, Rh and Pd) catalysts supported by thin active carbon cloth sheets and alumite plates \cite{kariya_efficient_2003}. A set of thermophysical and thermochemical properties of  commercially available benzyltoluene and dibenzyltoluene and their hydrogenated derivatives were measured \cite{mueller_liquid_2015}. (De)hydrogenation between 1-methylindole and octahydro-1-methylindole were examined \cite{yang_study_2018} using such catalysts as Ru/Al$_2$O$_3$ and Pd/Al$_2$O$_3$. Among the LOHCs explored, \textit{N}-ethylcarbazole was proposed as a very promising candidate for its relatively low  $\Delta H^{\rm{dh}}$ and relatively high storage capacity (5.8 wt\%) \cite{modisha_prospect_2019}, and a number of studies \cite{sotoodeh_kinetics_2009, mehranfar_hydrogen_2015, eblagon_study_2010, sotoodeh_dehydrogenation_2012, fei_study_2017} have been carried out to explore its related thermodynamic and kinetic properties. For example, combined experimental and theoretical studies investigated the effects of terrace sites of Ru catalyst surface \cite{eblagon_study_2010} and Pd particle size \cite{sotoodeh_structure_2011} on (de)hydrogenation process of \textit{N}-ethylcarbazole. Nevertheless, unlike dibenzyltoluene and 1-methylindole which are liquid even below 0 $^\circ$C, \textit{N}-ethylcarbazole, with a melting point of 68 $^\circ$C, is in solid state at room temperature. Besides, the relatively weak bonding between the heteroatom (nitrogen) and the ethyl radical results in dealkylation \cite{gleichweit_dehydrogenation_2013} above ~120 $^\circ$C, although the heteroatom reduces the dehydrogenation enthalpy.

\begin{figure}
\includegraphics[width=3.3in]{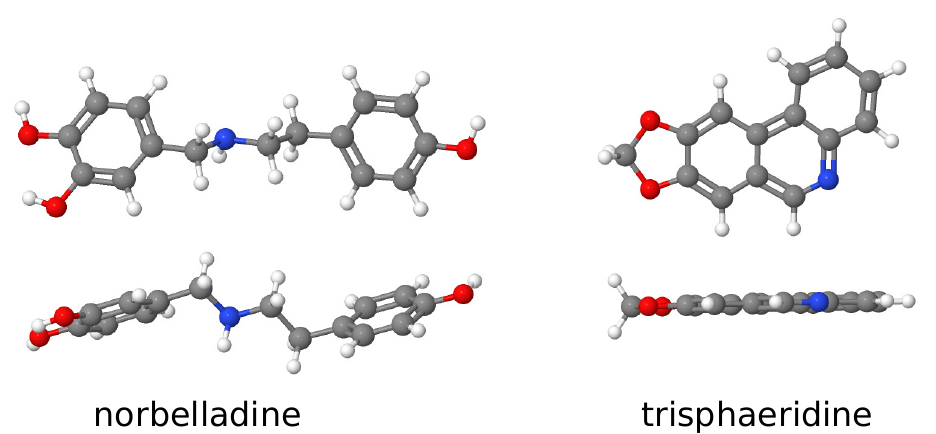}
\caption{Ball-stick schematics of norbelladine (C$_{15}$H$_{17}$NO$_3$) and trisphaeridine (C$_{14}$H$_{9}$NO$_2$). Color code: N (blue), O (red), C (grey), and H (white).}
\label{fig:struc}
\end{figure}

To date, searching for suitable LOHC systems is an important research field for hydrogen economy. The typical LOHCs, except $N$-ethylcarbazole, have high dehydrogenation enthalpy $\Delta H^{\rm{dh}}$ \cite{modisha_prospect_2019} and hence it's desirable to find more LOHC systems with relatively low $\Delta H^{\rm{dh}}$. Also, production of LOHCs from fossil-based materials is not green or sustainable, and manufacturing LOHC molecules using natural products is a promising alternative. On this regard, natural molecules out of plants represent an unexplored mine of possible LOHC candidates. Driven by these needs, in this work we explore the possibility of using Amaryllidaceae alkaloids \cite{reis_amaryllidaceae_2019}, which can be extracted from the family of Amaryllidaceae plants, as LOHCs. Specifically, we study the thermodynamics related to dehydrogenation of perhydro-norbelladine and perhydro-trisphaeridine based on first principles computations. Among the major Amaryllidaceae alkaloids \cite{li_amaryllidaceae_2020}, norbelladine and trisphaeridine (Fig. \ref{fig:struc}) have low molecular weights and high hydrogen storage capacity (4\% and 5.91\%, respectively). Our work indicates that trisphaeridine is a potential LOHC for its low (de)hydrogenation enthalpy while norbelladine has low (de)hydrogenation reversibility due to the stabler perhydro-norbelladine isomers related to dehydration. This work represents an initial demonstration of the rich potential of bio-based molecules for hydrogen storage. 

\begin{figure*}
\includegraphics[width=3in]{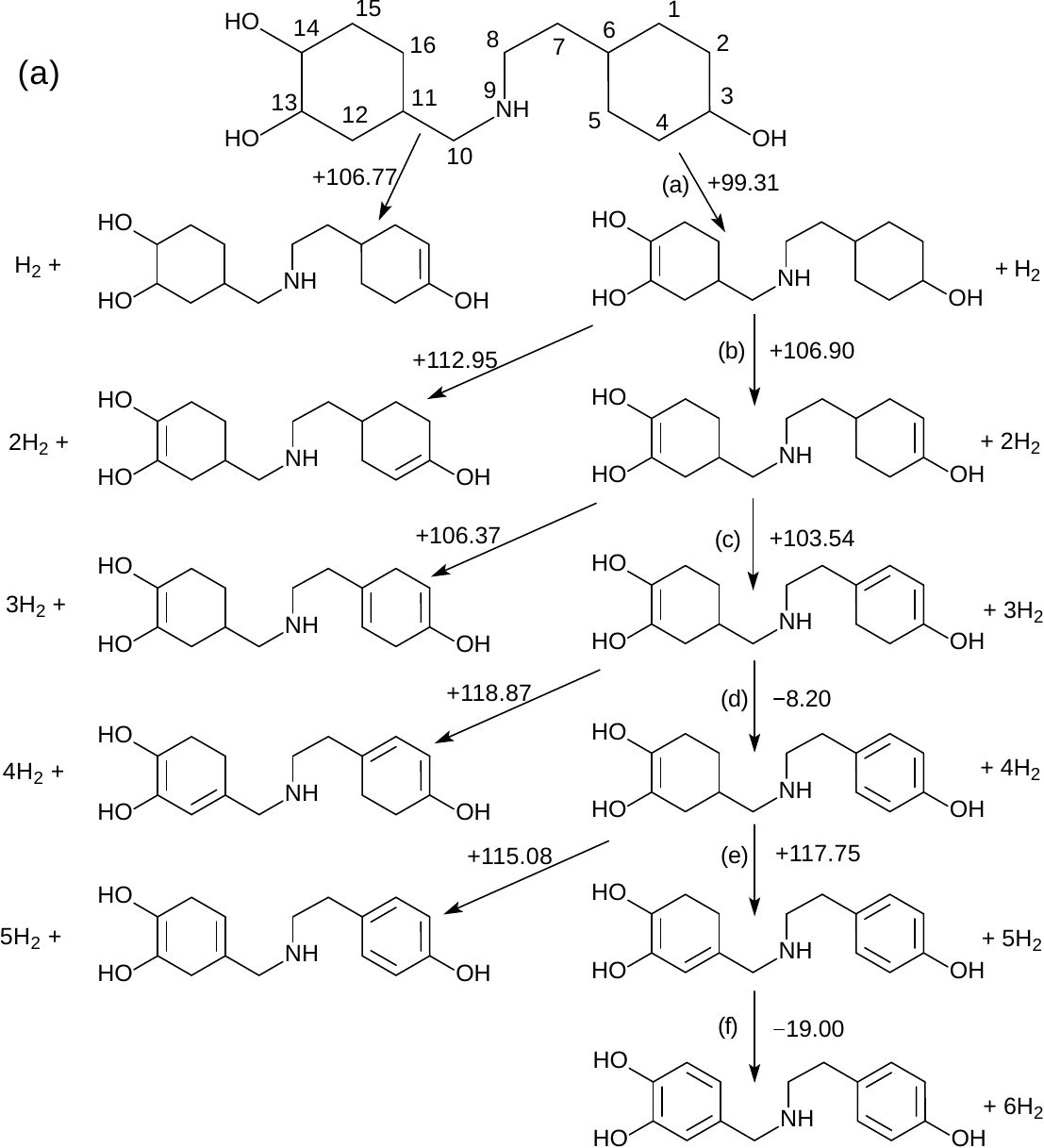}
\hspace{20pt}
\begin{minipage}{0.5\textwidth}
\vspace{-8 cm}
    \includegraphics[width=3in]{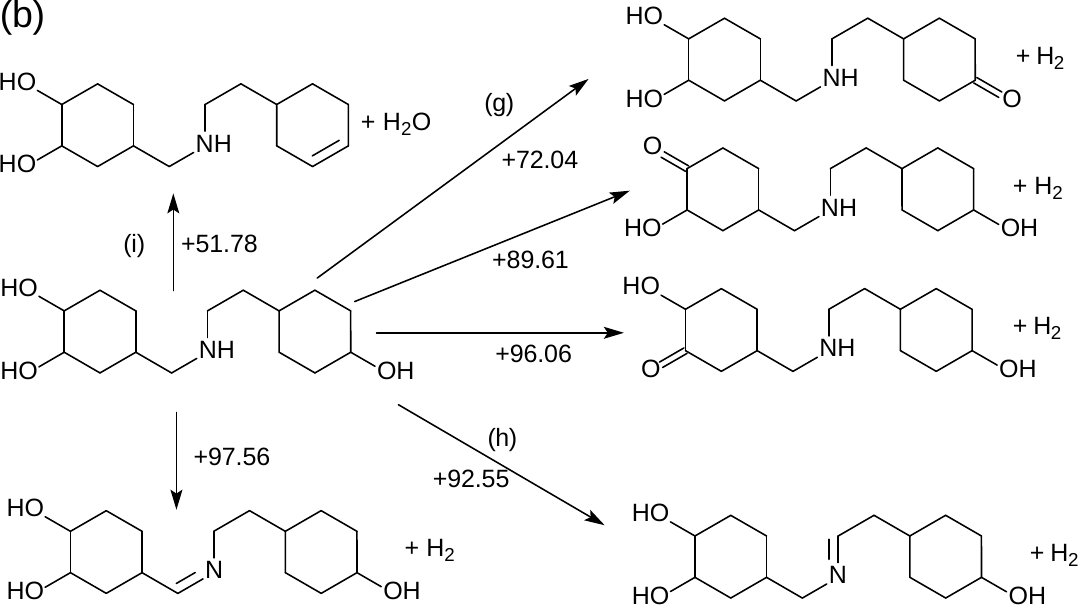}
\end{minipage}
\vspace{.5 cm}
\includegraphics[width=6.in]{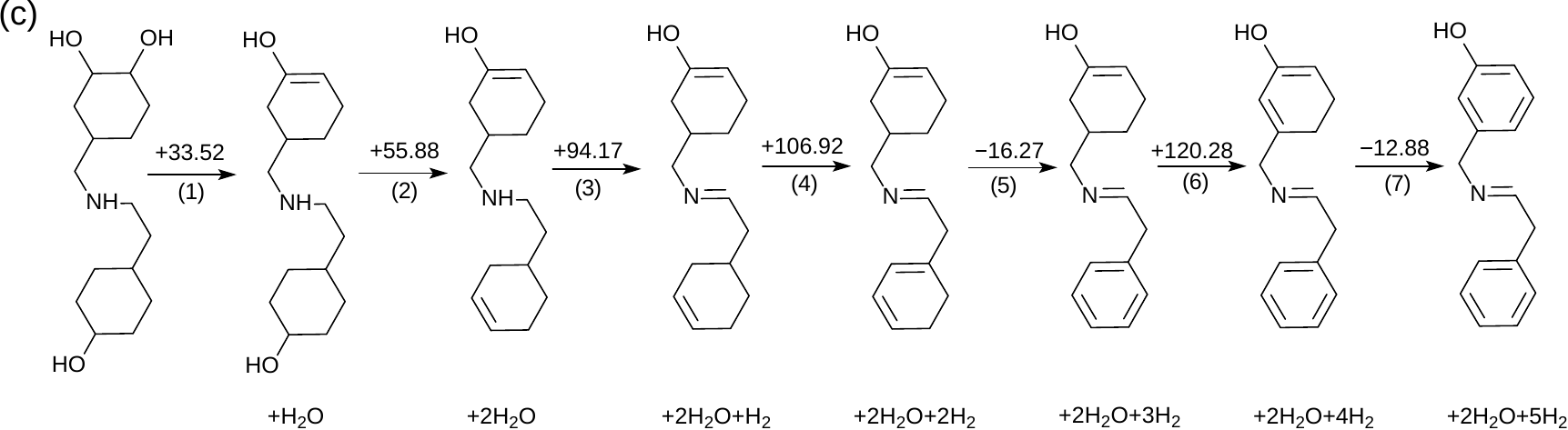}
\caption{Dehydrogenation of perhydro-norbelladine. (a) Step-wise dehydrogenation constrained on the carbon rings. (b) Some possible initial dehydrogenation or dehydration sites other than the rings. (c) Step-wise dehydrogenation/dehydration for the whole system. Energies are enthalpy change at 298 K and 1 atm, in KJ/mol-H$_2$ or KJ/mol-H$_2$O. The text like `+$n$H$_2$' indicates products of dehydrogenation.}
\label{fig:norbelladine}
\end{figure*}

\section{Methods}

Geometry optimization of norbelladine and trisphaeridine molecules, with a various number of hydrogen atoms bonded, was performed using the code VASP \cite{kresse_efficient_1996} with the GGA-PBE exchange-correlation functional and a plane wave basis set with an energy cutoff of 400 eV. For geometry optimization a simulation cell separating the molecules from its periodic images by more than 10 $ $\AA$ $ was used, and the $k$-point sampling was performed at the $\Gamma$ point. 

Energy cost for step-wise dehydrogenation, $\Delta E$, of perhydro-norbelladine and perhydro-trisphaeridine can be defined through equations (1) and (2), respectively:

\begin{align}
{\rm{C_{15}H}}_{17+i}{\rm{NO_3}} + \Delta E ={\rm{C_{15}H}}_{17+i-2}\rm{NO_3} +\rm{H_2}\\
{\rm{C_{14}H}}_{9+i}{\rm{NO_2}} + \Delta E ={\rm{C_{14}H}}_{9+i-2}\rm{NO_2} +\rm{H_2}
\label{eq:nuc}
\end{align}
where $i$ is the number of hydrogen atoms adsorbed. The average dehydrogenation energy cost per H$_2$ can be defined as $\sum_j \Delta E_j/n$, where $j$ indicates the steps and $n$ is the number of H$_2$ released along the path. This definition is equivalent to the dehydrogenation energy defined in the literature \cite{yang_study_2018} as $(E_{\rm{product}}+n\times E_{\rm{H_2}}-E_{\rm{reactant}})/n$, where the subscripts `reactant' and `product' represent hydrogenated and dehydrogenated molecules, respectively. Energy for reactions involving dehydration can be defined similarly. $\Delta E$ was computed at three levels of approximation: (1) DFT level (with HSE03 hybrid functional \cite{heyd_hybrid_2006} correction) ; (2) DFT with zero point energy correction ($E_{\rm{DFT}}^{\rm{ZPE}}$); and (3) $E_{\rm{DFT}}^{\rm{ZPE}}$ plus thermal correction (which gives enthalpy change $\Delta H$ and Gibbs free energy change $\Delta G$) where the molecules were treated as ideal gas at 298 K and 1 atm. During the thermal correction, the system volume is not specified as it is an implicit variable coupled with pressure. Zero point energy of the relevant molecules was obtained via calculating the vibrational frequencies, and thermal correction was performed using code VASPKIT \cite{wang_vaspkit_2019}. For thermal correction, the number of $\pi$-bonds of the molecules was counted for the lone electrons, which contribute to the electronic component of entropy.

\begin{figure*}
\includegraphics[width=6.in]{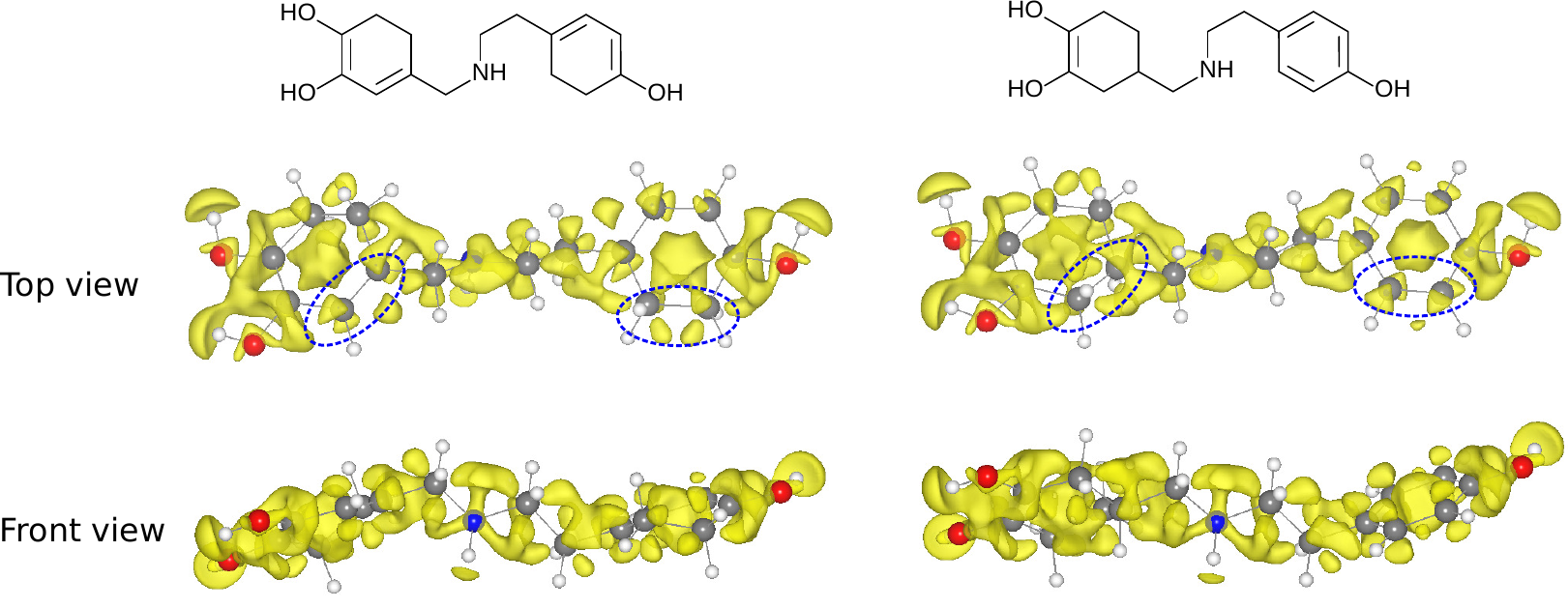}
\caption{Isosurface of differential charge density, or the charge density minus the superposition of atomic charge densities, for two partially dehydrogenated norbelladine conformations. As highlighted by the dashed ellipses, the formation of a double bond due to dehydrogenation enhances charge density around the double-bond C atoms in the direction normal to the aromatic ring. For the sextet conformation, this enables the delocalization of $\pi$-electrons over the aromatic ring. The isosurface represents charge density level of 0.5 $\mu e /$\AA$^3 $.}
\label{fig:norb-sextet}
\end{figure*}

\begin{figure}
\includegraphics[width=3.1in]{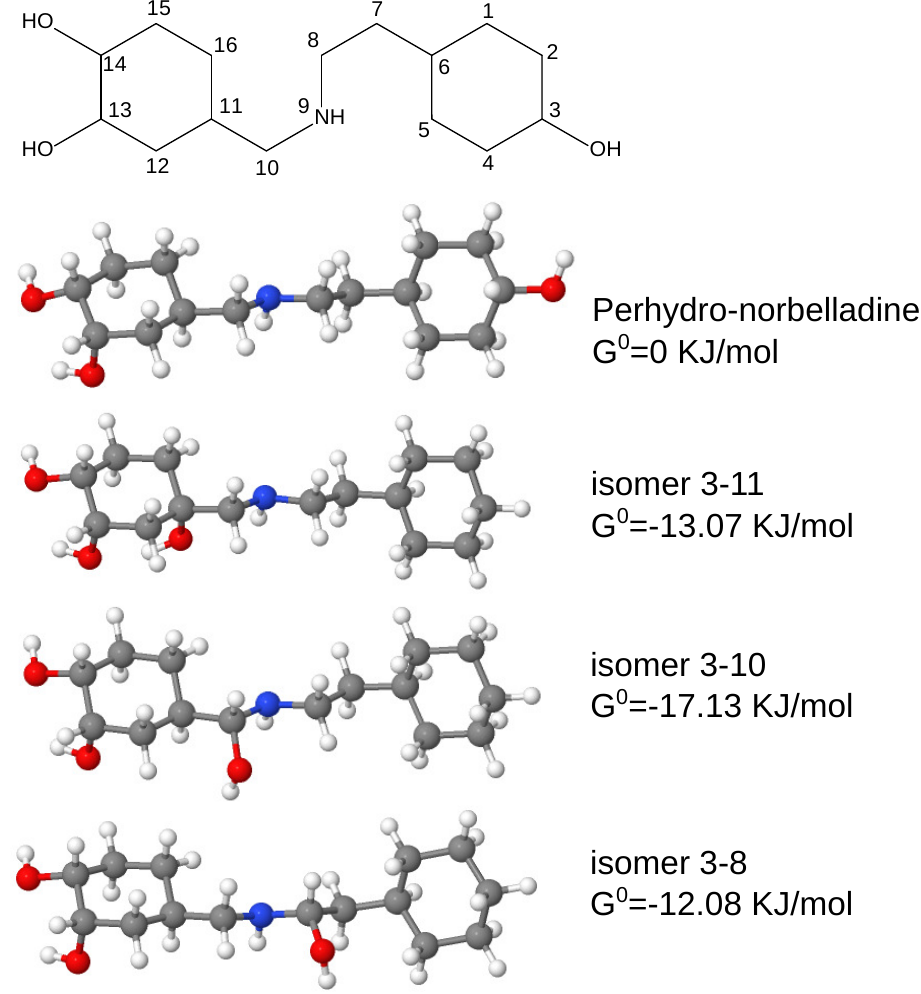}
\caption{ The schematic of perhydro-norbelladine and the relative standard Gibbs free energies of several example isomers. Term 3-11, for example, means the OH radical attached to site 3 is relocated to site 11.}
\label{fig:hydrogenation}
\end{figure}

\begin{figure}
\includegraphics[width=3.in]{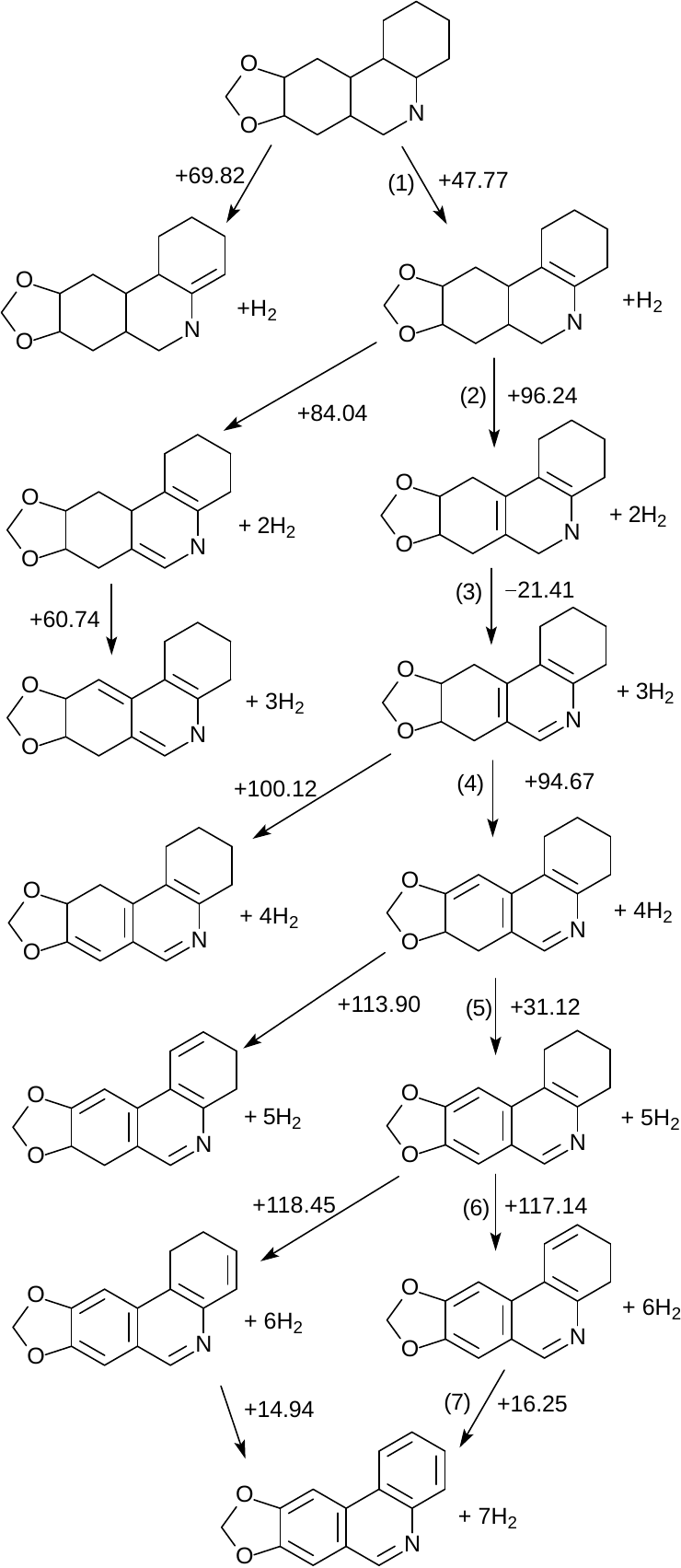}
\caption{ Dehydrogenation of perhydro-trisphaeridine. The arrows with step numbers indicate the favorable reaction path and the float numbers represent enthalpy change for each step at 298 K and 1 atm, in KJ/mol-H$_2$.}
\label{fig:trisphaeridine}
\end{figure}

\section{Results and discussions}
The dehydrogenation of perhydro-norbelladine is a relatively complex process due to the presence of the OH radicals and the N-doped carbon chain. We first considered the discharge of H$_2$ from the two cyclic rings (Fig. \ref{fig:norbelladine}(a)) in view that this is relatively simple, and the cyclic rings are representative of LOHC structures. In this case we considered the paired C atoms of each cyclic C-C bond as possible dehydrogenation sites. In the first step (as labelled (a) in the figure), the most favorable pair is sites 13-14 between the two hydroxyl radicals, followed by sites 2-3 on the other cyclic ring, with the standard enthalpy change $\Delta H^{\circ}$ being 99.31 and 106.77 KJ/mol-H$_2$, respectively. Dehydrogenation successively occurs on these two site pairs in steps (a) and (b), and, because of their far distance, the presence of $\pi$-bonding on sites 13-14 in step (a) hardly affect  $\Delta H^{\circ}$ for dehydrogenation on sites 2-3 in step (b). Upon the discharge of the third H$_2$ on sites 1-6, the following dehydrogenation is strongly favorable on sites 4-5, with negative (exothermic) $\Delta H^{\circ}$ being $-$8.2 kJ/mol-H$_2$, because this promotes the formation of an aromatic sextet. In comparison, $\Delta H^{\circ}$ for dehydrogenation at the next favorable sites, 11-12, is 118.87 KJ/mol-H$_2$. The formation of aromatic sextet enhances the charge density of $\pi$-electrons over the ring (Fig. \ref{fig:norb-sextet}) and shrinks the C-C distance from 1.53 $ $\AA$ $ for the fully hydrogenated ring to 1.40 $ $\AA$ $, resulting in the delocalization of $\pi$-electrons over the ring. Our calculations in the first step show that $\Delta H^{\circ}$ for dehydrogenation on sites 4-5 alone is 135.42 KJ/mol-H$_2$, and hence the negative enthalpy change in step (d) associated with this pair of sites exemplifies the impact of adjacent $\pi$-bonds on the thermodynamics of dehydrogenation. Sites 11-12 are the next most favorable position in step (d) and also the most favorable in the following step (step (e)). The value of $\Delta H^{\circ}$ for sites 11-12 in step (d), 118.87 kJ/mol-H$_2$, is very close to step (e), 117.75 kJ/mol-H$_2$, suggesting that dehydrogenation at sites 11-12 is not affected by the formation of the sextet in the other cyclic ring because of the far distance. Similar phenomenon was observed in step (b), where dehydrogenation at sites 2-3 is independent of the previous hydrogen release at far sites 13-14.  Discharge of the last H$_2$ in step (e) is exothermic ($-$19 KJ/mol-H$_2$), in favor of sextet formation. On average, the release of hydrogen from the polycyclic rings occurs with an enthalpy cost of 66.72 KJ/mol-H$_2$.

For perhydro-norbelladine, discharge of hydrogen from sites other than the cyclic rings can be more energetically favorable. As shown in Fig. \ref{fig:norbelladine}(b), the release of two H atoms from a hydroxyl radical and the neighboring C atom, thus forming a carbonyl group and a discharged H$_2$, corresponds to $\Delta H^{\circ}$ from $\sim$72 to $\sim$ 96 KJ/mol-H$_2$. Enthalpy change for releasing H atoms from the nitrogen site and its neighboring C site is around 92-97 KJ/mol-H$_2$. Even lower $\Delta H^{\circ}$, ranging from $\sim$34 to $\sim$64 KJ/mol-H$_2$, was found for dehydration of perhydro-norbelladine where a hydroxyl radical and  a near H atom are released. Production of H$_2$O in LOHC dehydrogenation could be beneficial for the thermodynamic process. It has been reported that the production of H$_2$O via oxidation of the hydroxymethyl group is exothermic ($-$188 KJ/mole) and, hence, promotes autothermal dehydrogenation of Perhydro-3-Hydroxymethyl-\textit{N}-Methylcarbazole \cite{cooper_hydrogen_2007}.

\begin{table*}
\caption{Computed energies for marked dehydrogenation steps of perhydro-norbelladine and perhydro-trisphaeridine in Fig. \ref{fig:norbelladine}(c) and Fig. \ref{fig:trisphaeridine}. Standard conditions (298 K and 1 atm) were used for computing $\Delta H$ and $\Delta G$. Energy is in KJ/mol-H$_2$ and, for steps 1 and 2 of perhydro-norbelladine, KJ/mol-H$_2$O.}\tabcolsep=8pt
\begin{tabular} {@{\extracolsep{4pt}}ccccccccc}  \hline  \hline
Reaction step & \multicolumn{4}{c}{Perhydro-norbelladine}  &  \multicolumn{4}{c}{Perhydro-trisphaeridine}  \\
~ & $\Delta E_{\rm{DFT}}$ & $\Delta E_{\rm{DFT}}^{\rm{ZPE}}$ & $\Delta H^{\circ}$ & $\Delta G^{\circ}$ & $\Delta E_{\rm{DFT}}$ & $\Delta E_{\rm{DFT}}^{\rm{ZPE}}$ & $\Delta H^{\circ}$ & $\Delta G^{\circ}$\\
\cline{1-5}
\cline{6-9}
1 & 45.29 & 27.74 & 33.52 & -16.01  & 75.53 & 39.48 & 47.77 & 7.64 \\
2 & 62.53 & 45.13 & 50.88 & 3.89  & 124.84 & 86.63 & 96.24 & 52.18 \\
3 & 123.25 & 86.45 & 94.17 & 55.09  & 3.46 & -26.95 & -21.41 & -52.67 \\
4 & 135.21 & 98.83 & 106.92 & 66.86 & 121.72 & 86.86 & 94.67 & 57.58 \\
5 & 5.43 & -22.80 & -16.27 & -51.35  & 56.93 & 23.05 & 31.12 & -7.73 \\
6 & 148.19 & 112.31 & 120.28 & 82.95  & 144.84 & 109.12 & 117.14 & 78.33 \\
7 & 10.82 & -20.09 & -12.88 & -54.54 & 40.44 & 9.44 & 16.25 & -19.11 \\
\hline\hline
\end{tabular}
\label{tab:steps}
\end{table*}
\vspace{20pt}

Equipped with the above understanding we studied the feasible dehydrogenation path, as shown in Fig. \ref{fig:norbelladine}(c). In addition to the formation of sextets, it features the production of 2H$_2$O, one from each cyclic ring, and a $\pi$-bond between N and its neighboring C. Similar to the constrained dehydrogenation process on the rings, the two steps here corresponding to sextet formation are exothermic and hence represent the intermediate and final products, respectively. We denote the steps along this path as (1)-(7) to differentiate them from the steps mentioned above.

The removal of OH radicals is likely to reduce dehydrogenation reversibility of norbelladine. During the hydrogenation process, depending on a series of factors such as thermodynamics and the local concentrations of H cations and OH radicals, the original OH sites may be occupied by H cations and/or the OH radicals may be attracted to other sites. In view of the vast number of possible reactions involved, it is challenging to exhaustively explore all the intermediate and final structures. Here we examine (de)hydrogenation reversibility of norbelladine by comparing the stability of perhydro-norbelladine and a small subset of its isomers. Specifically, we fixed the two neighboring OH radicals at their original positions and switched the third OH radical with a hydrogen ion at various locations. Fig. \ref{fig:hydrogenation} shows three isomers, as examples, that are stabler than perhydro-norbelladine by more than 10 KJ/mol. This suggests the low reversibility of norbelladine as a LOHC candidate.

\begin{table*}
\begin{threeparttable}
\caption{Average dehydrogenation energy (KJ/mol-H$_2$) cost for example LOHC systems. Note that for perhydro-norbelladine only 5 H$_2$ were produced in the dehydrogenation process. The energies are DFT energy ($E$) with and without zero point energy correction, enthalpy ($H$), and Gibbs free energy ($G$). Subscript `expt' for experimental values. All values are from this work unless referenced.}\tabcolsep=9pt
\begin{tabular} {lllllll}  \hline  \hline

System & $\Delta E_{\rm{DFT}}$ & $\Delta E_{\rm{DFT}}^{\rm{ZPE}}$ & $\Delta H^{\circ}$ & $\Delta G^{\circ}$ & $\Delta H^{\circ}_{\rm{expt}}$ & $\Delta G^{\circ}_{\rm{expt}}$\\
\hline
Perhydro-norbelladine (C$_{15}$H$_{29}$NO$_3$) & 106.14 & 65.51 & 75.33 & 17.38 & ~ & ~\\
Perhydro-trisphaeridine (C$_{14}$H$_{23}$NO$_2$) & 81.11 & 46.81 & 54.54 & 16.60 & ~& ~\\
Cyclohexane (C$_6$H$_{12}$) & 92.29 & 61.41 & 67.56 & 35.88 & 68.7 \cite{pez_hydrogen_2006,mueller_thermodynamic_2013}& 32.6 \cite{mueller_thermodynamic_2013}\\
~ & ~ & ~ & 76.0 \cite{pez_hydrogen_2006} & ~ & ~& ~ \\
$Trans$-decalin (C$_{10}$H$_{18}$) & 97.88 & 63.35 & 70.93 & 33.32 & 66.3\cite{mueller_thermodynamic_2013} \tnote{*}& 20.5\cite{mueller_thermodynamic_2013} \tnote{*} \\
~ & ~ & ~ & 71.3 \cite{pez_hydrogen_2006} & ~ & 66.8 \cite{pez_hydrogen_2006}& ~\\
Perhydro-\textit{N}-ethylcarbazole (C$_{14}$H$_{25}$N) & 77.82 & 44.30 & 52.02 & 13.62  & 53.2 \cite{mueller_thermodynamic_2013, verevkin_thermodynamic_2011}& 18.2 \cite{mueller_thermodynamic_2013}\\
~ & ~ & ~ & ~ & 3.42 \cite{mehranfar_hydrogen_2015} & ~& ~ \\
Methylcyclohexane (C$_7$H$_{14}$) & 100.85 & 67.08 & 74.82 & 35.87 & 68.3 \cite{mueller_thermodynamic_2013}& 31.6 \cite{mueller_thermodynamic_2013}\\
Octahydro-1-methylindole (C$_9$H$_{17}$N) & 85.99 & 52.43 & 60.19 & 21.53 & ~& ~\\
~ & 79.6\cite{yang_study_2018} \tnote{\dag}& ~ & ~ & ~ & ~& ~ \\
\hline\hline
\end{tabular}
\begin{tablenotes}\footnotesize
\item[*] The state ($trans$- or $cis$-) of decalin is not specified in the reference. Usually the two states of decalin can coexist (see text).
\item[${\dag}$] Values from standard density functional theory calculations using PW91 exchange-correlation functional without hybrid functional correction. Correspondingly, our standard density functional theory calculations using PBE exchange-correlation gives $\Delta E$ of 71.76 kJ/mol-H$_2$.
\end{tablenotes}
\label{tab:enthalpy}
\end{threeparttable}
\end{table*}
\vspace{20pt}
 
Next, we study the dehydrogenation of perhydro-trisphaeridine. As shown in Fig. \ref{fig:trisphaeridine}, $\Delta H^{\circ}$ for the most favorable initial dehydrogenation sites, at the fused C-C pair of the end cyclic carbon ring, is 47.77 KJ/mol-H$_2$, or $\sim$22 KJ/mol-H$_2$ lower than that of the next favorable sites. For step 2, the most favorable sites, with $\Delta H^{\circ}$ being 84.04 KJ/mol-H$_2$, are the C-C pair in the nitrogen-containing ring that is parallel to the dehydrogenated one in step 1. However, this hampers the formation of sextet in the nitrogen-containing ring and results in relatively high $\Delta H^{\circ}$ (60.74 KJ/mol-H$_2$) in the following step. On the other hand, the second most favorable sites in step 2 are the other fused C-C pair of the nitrogen-containing ring. This triggers the exothermic sextet formation of the nitrogen-containing ring in step 3 and turns out to be the overall most favorable path. In the following steps 4 and 5, a sextet forms out of the ring next to the oxygen-containing five-membered ring, followed by sextet formation of the end cyclic carbon ring in steps 6 and 7. It is interesting to note that steps 5 and 7, which form the sextets, are endothermic, as opposed to the exothermic formation of the nitrogen-containing sextet. This suggests that heteroatoms help reduce the enthalpy change of dehydrogenation \cite{pez_hydrogen_2006}.

In Table \ref{tab:steps} we list the computed DFT energies with various levels of correction for the above-mentioned reaction steps. Among the corrections, the zero point energy and entropy affect the thermodynamics significantly. It is also worth to note that all the reaction steps that close a sextet (i.e., steps 5 and 7 for norbelladine and steps 3, 5, and 7 for trisphaeridine) spontaneously occur because of their negative Gibbs free energy change.

As the study on Amaryllidaceae alkaloids for LOHC applications is new, relevant experimental data are scarce. To validates our calculations, we have also computed the dehydrogenation energetics of some typical LOHCs with existing experimental data. Table \ref{tab:enthalpy} compares the average dehydrogenation energies of perhydro-norbelladine and perhydro-trisphaeridine with some commonly studied LOHC candidates. As can be seen for the typical LOHCs, including cyclohexane, $trans$-decalin, perhydro-$N$-ethylcarbazole, and methylcyclohexane, the computed $\Delta H^{\circ}$ values compare reasonably well with the available experimental values. Specifically for cyclohexane and perhydro-$N$-ethylcarbazole, the error for calculated standard enthalpy changes with respect to their experimental values are within 3\%, and the error for $trans$-decalin and methylcyclohexane are within 10\%. For the calculated Gibbs free energy $\Delta G^{\circ}$, the error ranges from 10\% to 25\% except for $trans$-decalin, for which the calculated $\Delta G^{\circ}$ is about 63\% higher than the experimental value. Although $trans$-decalin is the thermodynamically stable state, upon the hydrogenation of naphthalene the less stable $cis$-decalin is found to coexist with $trans$-decalin in a ratio depending on the processing \cite{park_characteristics_2002, albertazzi_hydrogenation_2003}. The existence of $cis$-decalin results in a reduced dehydrogenation enthalpy change with respect to pure $trans$-decalin, and in turn a reduced $\Delta G^{\circ}$. Comparing with other first principles calculations, we noted our results are closer to the experimental values. For example, for $\Delta H^{\circ}$ of cyclohexane (experimental value 68.7 kJ/mol-H$_2$), our and literature \cite{pez_hydrogen_2006} values are 67.6 and 76.0 kJ/mol-H$_2$, respectively, and for $\Delta G^{\circ}$ of perhydro-$N$-ethylcarbazole (experimental value 18.2 kJ/mol-H$_2$), our and literature \cite{mehranfar_hydrogen_2015} values are 13.6 and 3.4 kJ/mol-H$_2$, respectively. As can be seen, overall our calculations for existing LOHCs match well with experiments, which validates the methods for this study.  

\begin{figure}
\includegraphics[width=3.6in]{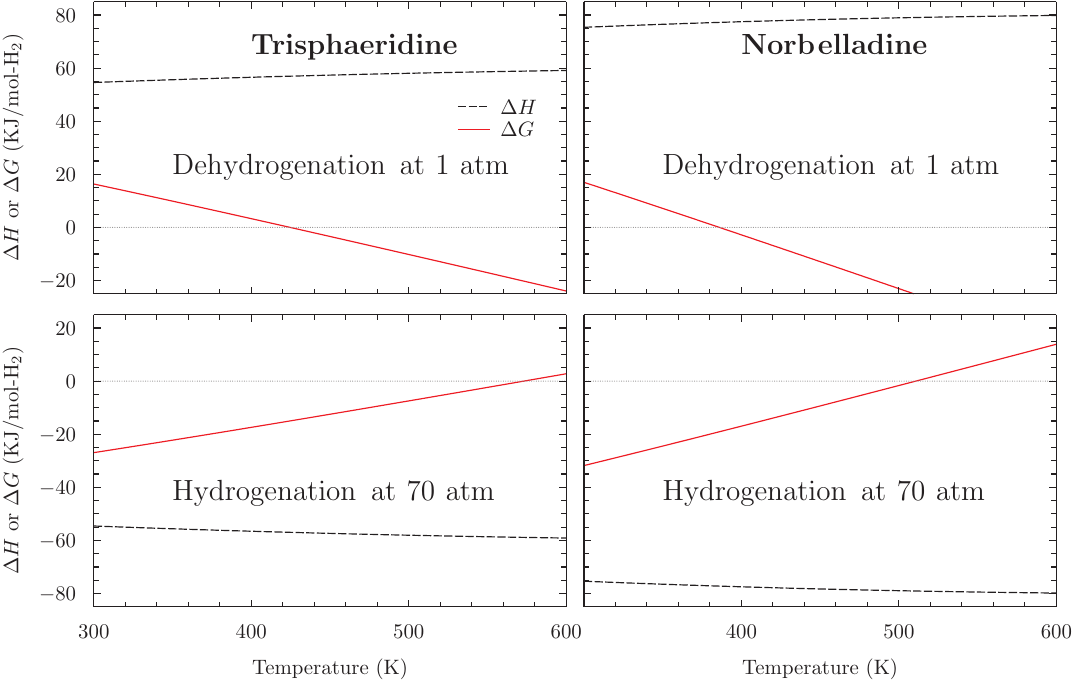}
\caption{Temperature dependence of enthalpy and Gibbs free energy changes for (de)hydrogenation of trisphaeridine (left columns) and norbelladine (right columns).}
\label{fig:temp-dh-dg}
\end{figure}

Notably, the overall enthalpy change for dehydrogenation of perhydro-trisphaeridine is similar to that of perhydro-\textit{N}-ethylcarbazole and lower than those of the other systems. Based on thermal correction using varying temperatures and pressures, we evaluated the energy changes for (de)hydrogenation of (perhydro-)trisphaeridine and (perhydro-)norbelladine as a function of temperature. As can be seen from Fig. \ref{fig:temp-dh-dg}, from 300 to 600 K, the enthalpy change $\Delta H$ for dehydrogenation at 1 atm increases from $\sim$55 KJ/mol-H$_2$ by $\sim$10\% only, and $\Delta G$ constantly decreases, crossing the $\Delta G$=0 line at $\sim$430 K. Hydrogenation at 70 atm, on the other hand, is exothermic within the whole temperature range, and $\Delta G$ is less than zero for temperature below 570 K. Similar trends were found for (perhydro-)norbelladine. The above discussions suggest, from the perspective of energetics, the potential of trisphaeridine as an ideal LOHC candidate. We note that, nevertheless, experimental data for other important properties, such as melting/boiling points, vapour pressure, thermal stability, toxicity, and catalytic (de)hydrogenation kinetics, are still scarce for trisphaeridine and other Amaryllidaceae alkaloids. These properties need to be investigated for final evaluation of a molecule for LOHC applications and they also represent new research opportunities in this field. From the structural perspective we expect trisphaeridine to have better thermal stability than $N$-ethylcarbazole because it does not contain the alkyl tail that could result in dealkylation at high temperatures as in $N$-ethylcarbazole \cite{gleichweit_dehydrogenation_2013}. Also, used as medicine molecules, Amaryllidaceae alkaloids should be generally nontoxic. From the calculation perspective, we noted some recent efforts in predicting melting points of organic molecules based on machine learning \cite{mcdonagh_predicting_2015}. Such methods are valuable for large scale investigation of Amaryllidaceae alkaloids and other molecules for LOHC applications in the future.

\section{Summary}
In this work we explored the possibility of using natural-products-based molecules of norbelladine and trisphaeridine, two Amaryllidaceae alkaloids, as liquid organic hydrogen carriers (LOHCs) by studying dehydrogenation thermodynamics of perhydro-norbelladine and perhydro-trisphaeridine based on first principles computations. It was found that for norbelladine the energetically favorable reaction path produces 2 H$_2$O and 5 H$_2$ with an average enthalpy change of $\sim$75 KJ/mol-H$_2$ at the standard condition.  Upon hydrogenation, however, stabler isomers of perhydro-norbelladine were found, which suggests the low reversibility of (perhydro-)norbelladine. 

For perhydro-trisphaeridine, the dehydrogenation process produces 7 H$_2$ with an average standard enthalpy change of $\sim$54 KJ/mol-H$_2$, similar to that of perhydro-\textit{N}-ethylcarbazole, a typical LOHC known for its low dehydrogenation enthalpy change. At temperature ranging from 300 to 600 K, the enthalpy change is relatively stable and the Gibbs free energy change ($\Delta G$) decreases from $\sim$17 to $\sim-$24 KJ/mol-H$_2$, with $\Delta G$ being zero at $\sim$420 K. For hydrogenation at a pressure of 70 atm, $\Delta G$ increases with temperature but is negative at temperatures below 570 K. The favorable thermodynamics of trisphaeridine, along with its relatively high hydrogen storage capacity ($\sim$5.9 wt\%), suggests its promising potential as a liquid organic hydrogen carrier. In view that typical LOHCs except perhydro-\textit{N}-ethylcarbazole have high dehydrogenation enthalpy and LOHCs based on fossil materials are not sustainable, the finding of bio-based trisphaeridine with low dehydrogenation enthalpy represents a new direction for searching suitable LOHCs in the future. Of course, for its real LOHC application, other properties, such as melting/boiling temperatures and its catalytic (de)hydrogenation kinetics, still need evaluation.

There are about 500 known Amaryllidaceae alkaloids \cite{li_amaryllidaceae_2020} and the possibility of finding LOHCs, which may be better than trisphaeridine, can be further extended using the approach in this work.

\vspace{10pt}
\noindent{\bf{Acknowledgments}}

CT thanks the financical support from the Australian National University Grand Challenge program (Zero-Carbon Enerygy for the Asia-Pacific). SF thanks the financical support from the National Natural Science Foundation of China (No. 21802129).
We acknowledge the access to the structural database constructed from the repository and collection of 600 samples of Amaryllidaceae alkaloids provided by Dr Henry M. Fales, NHLBI Laboratory of Applied Mass Spectrometry,  National Institutes of Health, Bethesda, to Professor Martin G. Banwell, Research School of Chemistry, Australian National University, Canberra, for selecting trisphaeridine and norbelladine for this work.
This research was undertaken with the assistance of resources from the National Computational Infrastructure (NCI Australia), an NCRIS enabled capability supported by the Australian Government.

\vspace{10pt}
\noindent{\bf{Reference}}

\end{document}